\documentclass[aps,superscriptaddress,showpacs,nofootinbib,eqsecnum,prd,notitlepage]{revtex4-1}
\pdfoutput=1

\usepackage{graphics} \usepackage{epsf} \usepackage{eucal}
\usepackage{amssymb} \usepackage{amsmath} \usepackage{graphicx}
\usepackage{bm} \usepackage{dcolumn} \usepackage{epsfig}
\usepackage{multirow} \usepackage{hyperref} \usepackage{subfigure}
\usepackage{appendix} \usepackage{flafter}

\newcommand{\be}{\begin{equation}} \newcommand{\ee}{\end{equation}}
\newcommand{\bea}{\begin{eqnarray}} \newcommand{\eea}{\end{eqnarray}}

\begin{document} 

\title{Theory of non-Gaussianity in warm inflation}

\author{Mar Bastero-Gil} \email{mbg@ugr.es} \affiliation{Departamento
  de F\'{\i}sica Te\'orica y del Cosmos, Universidad de Granada,
  Granada-18071, Spain}

\author{Arjun Berera} \email{ab@ph.ed.ac.uk} \affiliation{SUPA, School
  of Physics and Astronomy, University of Edinburgh, Edinburgh, EH9
  3JZ, United Kingdom}

\author{Ian G. Moss} \email{ian.moss@ncl.ac.uk} \affiliation{School of
  Mathematics and Statistics, Newcastlle University, NE1 7RU, United
  Kingdom}

\author{Rudnei O. Ramos} \email{rudnei@uerj.br}
\affiliation{Departamento de F\'{\i}sica Te\'orica, Universidade do
  Estado do Rio de Janeiro, 20550-013 Rio de Janeiro, RJ, Brazil}



\begin{abstract}

The theory and methodology is
developed to compute the bispectrum in warm inflation, leading to results
for the non-linearity parameter and the shape of the bispectrum.
Particular attention is paid to the study of the
bispectrum in  the regime of weak dissipation and how stochastic
fluctuations affect the bispectrum.  It is shown that, in contrast
to the strong dissipative regime, the
amplitude of non-Gaussianity is strongly dependent on the 
parameters governing the microscopic physics
in the intermediate and weak dissipation warm inflation regimes. The most
important results concern the shape of the bispectrum, which has two
different, but distinct, forms in the weak and strong dissipative regimes.

\end{abstract}


\maketitle
\section{Introduction}
\label{sec1}

Inflation remains one of the most appealing solutions to the
cosmological puzzles.  Observations from the cosmic microwave
background are consistent with density perturbations that are very
close to Gaussian and scale invariant.  Measurements on non-Gaussianty
from the first year Planck data~\cite{planckng} show that they are
severely constrained although there remains room that there could be a
detectable signal. 

There are two dynamical pictures of inflation that have been
developed.  In one the scalar inflaton field is pictured to be almost
non-interacting with all other fields~\cite{inflation}.  Thus, the
Universe inflates in a vacuum state and the evolution of the scalar
inflaton field is governed by zero temperature physics.  This is the
cold inflation picture.  In the alternative picture, particle
production occurs concurrent to inflationary expansion. The scalar
inflaton field is governed by fluctuation-dissipation
dynamics~\cite{bf2} that controls the seeds of density fluctuations.
This is the warm inflation picture~\cite{wi}.  In this picture, for
the simplest dynamics the seeds of density perturbation are
thermal~\cite{im,bf2,abnpb}. The interaction of the scalar field with
other fields leads to particle production, which then leads to a
dissipation and fluctuation term in the inflaton evolution
equation~\cite{abwi2,bgr1}. We should note that since its original proposal
warm inflation has evolved quite considerably. 
Originally, dissipation was computed in the high temperature regime
and it was soon realized that thermal corrections to the inflaton potential 
would spoil inflation~\cite{bgr1,Yokoyama:1998ju}. However, 
more recent model building realizations easily overcomes these early concerns 
about the implementation of the warm inflation idea (a detailed
discussion of these issues and their solution is particularly
discussed extensively, for example, in the review
paper~\cite{Berera:2008ar}). The basic idea is that dissipation is
driven by the coupling of the inflaton to massive fields (with a mass
larger than the temperature of the thermal bath), which in turn
couple to light (relativistic) degrees of freedom. Therefore, in this
regime thermal corrections to the effective potential are  not an issue.

Recent findings reported by the BICEP2
collaboration~\cite{Ade:2014xna}  of a possible primordial tensor mode
signal lead to a possible additional observable to help discriminate
different inflation models as well as between these two paradigms of
inflation.  However, this alone may not be sufficient, since both
paradigms have promising models for explaining a tensor mode at
various energy scales.  The most promising hope for discriminating
between the two inflationary paradigms could come from measurement of
non-Gaussianity.  
As the measurements of the
cosmic microwave background  (CMB) radiation becomes more and more
precise, it is expected that not only will the
magnitude for the non-Gaussianity be measured, 
but also be possible to determine its shape.
In fact, more important than the magnitude of this effect
would be the shape of the bispectrum.
Large classes of inflation models can be described by different
bispectrum shapes, as discussed in details in
Ref.~\cite{Fergusson:2008ra}.  Among the various possibilities, warm
inflation has its distinctive ``warm"
shape~\cite{Moss:2007cv}, which is very  different from other more
common shapes, like the equilateral, local, flat and others. In
this paper we will develop the theory of non-Gaussianity in warm
inflation building on the previous works starting with the first
analysis of effects from the inflaton evolution equation~\cite{Gupta:2002kn}, 
followed by an analysis of the general relativity
perturbation equations~\cite{Moss:2007cv}.  The latter paper in
particular developed the theory for the regime where the dissipative
coefficient is bigger than the Hubble scale, which has been referred to
as strong dissipative warm inflation. In this paper we will develop
the corresponding theory for when the dissipative coefficient is
smaller than the Hubble scale, which is called weak dissipative warm
inflation.  Our work will also further develop the theory for the
strong dissipative regime.  In particular, a full analysis was recently
done on all stochastic forces present during warm inflation
and their effects on first-order cosmological perturbations~\cite{Bastero-Gil:2014jsa}.  
In this paper we will look at all
the second-order effects from all these sources of stochastic forces
and, thus, their effect on non-Gaussianity in both the weak and strong
warm inflation regimes.
 
Our results might be of importance in other
contexts as well, such as non-Gaussianity in curvaton models and in those
models where the curvature perturbations and the non-linearity is
dependent on the details of the reheating dynamicss.
In addition any inflation scenarios involving
particle production will in general
have some form
of backreaction dissipative effects, which necessarily
will be accompanied also
with fluctuation forces.  All such scenarios basically follow
the warm inflation picture.
The details of these effects associated with the particle production
can vary, but the
general approach adopted in this paper would also apply to such circumstances.

This paper is organized as follows. In section~\ref{sec2} we review the
fluctuations equations at first- and second-order for warm
inflation. In section~\ref{sec3} we present the scheme developed to compute
the bispectrum and the results for the non-linearity
parameter. In section~\ref{sec4} we study the shapes of the bispectrum for warm
inflation. We present our conclusion in section~\ref{sec5}.  Two appendices
are included where we show some of the technical details.

\section{Perturbations: First- and second-order}
\label{sec2}

Our aim is to analyse the non-Gaussianities generated by a mixture of a
slow-rolling scalar field and a radiation fluid during warm
inflation. We do this by constructing equations for a gauge invariant 
variable $\Phi(k,t)$ and evaluating the second order contribution to the
bispectrum,

\begin{equation}
B(k_1,k_2,k_3)\delta(\Sigma k)=\sum_{\rm cyc} \langle
\Phi_1(k_1,t_f)\Phi_1(k_2,t_f)\Phi_2(k_3,t_f)\rangle,
\end{equation}
where `cyc' denotes the set of cyclic permutations. The subscript in
$\Phi$ denotes first- and second-order perturbations, and the
perturbations are evaluated at a final time $t_f$ some e-folds after
horizon crossing.  We shall use the $\zeta$-variable \cite{Bardeen:1983qw}, 
which can be defined in terms of the total density perturbation $\delta\rho$ 
on surfaces of constant curvature by,

\begin{equation}
\Phi={\delta\rho\over 3(p+\rho)}.
\end{equation}
A large scale approximation, or the `delta N' approach, can be used to
show that this variable approaches a constant value on large
scales~\cite{Wands:2000dp,constantPhi}.

The second-order strategy follows previous work for bispectra in warm
inflationary models \cite{Moss:2011qc,Moss:2007cv,Moss:2007qd},  and
focuses on contributions to the non-Gaussianity which are of order one
in the slow-roll approximation. This makes it possible to discard many
terms in the perturbation equations that are similar in size to the
slow-roll parameters, in the way described below. We work in constant
curvature gauge, dropping the metric perturbations. This is justified
in the appendix~\ref{appA}, where we  show that the metric
perturbations are of the same order as the slow-roll parameters.

In warm inflation, the inflaton field is coupled to radiation during
inflation.  We consider the situation where the radiation is close to
thermal equilibrium, with temperature $T$ and four-velocity $u^a$.  A
covector $n_a$ is chosen orthogonal to the surfaces of constant time
$t$, with spatial coordinates $x^\alpha$ and spatial derivatives
$\partial_\alpha$.  The inflaton $\phi$ satisfies a stochastic
evolution equation with Gaussian noise term $\xi$,

\begin{equation}
\ddot\phi+3H\dot\phi+V_{,\phi}(\phi)+\Upsilon D\phi-\partial^2\phi=(2\Upsilon
T)^{1/2}\xi,
\label{infeq}
\end{equation}
where $H=\dot a/a$ is the expansion rate and
$\partial^2=a^{-2}\delta^{\alpha\beta}\partial_\alpha\partial_\beta$.
Dissipation effects are strongly influenced by the radiation, and they
are evaluated in the rest frame of the fluid. The dissipation depends
on a coefficient $\Upsilon(\phi,T)$ and the derivative along the fluid
four-vector $D\phi=u^a\nabla_a\phi$ (in our notation
latin indices mean space-time coordinates, while Greek ones refer to space 
components only).

Normal frame quantities will be used for the second-order theory.  The
`normal-frame' approach used below closely follows
ref.~\cite{Noh:2004bc}.  The fluid part of the stress-energy tensor is
expressed in the form

\begin{equation}
T^{r}_{ab}=(p_r+\rho_r)n_an_b+p_rg_{ab}+q_an_b+q_bn_a +\Pi_{ab}\,,
\end{equation}
where $q_a$ and $\Pi_{ab}$ are orthogonal to the normal direction. The
scalar perturbations of the fluid in the normal frame are defined by
the replacement 

\begin{eqnarray}
\rho_r&\to&\rho_r+\delta\rho^N\,,  \\ p_r&\to&p_r+\delta p^N\,,
\\ q_\alpha&\to&(1+w)\rho_r\delta v_\alpha^N,\qquad \delta
v_\alpha^N=\partial_\alpha \delta v^N\,.
\end{eqnarray}
These perturbations are expanded as a series, for example,
 
\begin{equation}
\delta\rho^N=\delta_1\rho^N+\delta_2\rho^N+\dots
\end{equation}

The rest frame of the radiation fluid is called the energy frame, and
the velocity perturbation in the energy frame is defined by

\begin{equation}
u_a=\gamma(n_a+\delta v_a^E),
\end{equation}
where $\gamma$ is the Lorentz factor. The density and pressure
perturbations in the energy frame are $\delta\rho^E$ and $\delta p^E$,
and we take the standard relation for a radiation fluid: $\delta
p^E=\delta\rho^E/3$.   The first-order perturbations are the same in
the energy or the normal frame,  and $\delta p^N=\delta\rho^N/3$, but
at second-order,

\begin{eqnarray}
\delta_2\rho^N&=&\delta_2\rho^E+
\frac43\rho_r(\delta_1v_\alpha^E)(\delta_1v^{E\alpha})\,,
\label{rhone}\\
\delta_2 v_\alpha^N&=&\delta_2 v_\alpha^E
+\frac{\delta_1\rho^E}{\rho_r}\delta_1v_\alpha^E\,,\\ 
\delta_2\Pi_{\alpha\beta}&=&\frac43\rho_r\left(
\delta_1v_\alpha^E \delta_1v_\beta^E
-\frac13h_{\alpha\beta}\delta_1v_\gamma^E \delta_1v^{E\gamma}
\right)\,.
\label{pi}
\end{eqnarray}

At the background level, the radiation fluid is sourced by the
inflaton field $\phi$ through a dissipation coefficient $\Upsilon$. In
the slow-roll approximation ($\dot \phi^2 \ll V(\phi)$, $\ddot \phi
\ll 3 H \dot \phi$), we have:

\bea  \dot \phi &\simeq& {-V_{,\phi}\over 3H(1+Q)}\,, \\ \rho_r &
\simeq& \frac34Q \dot \phi^2  \,,    \eea  where $Q=\Upsilon/(3H)$ and
again we have taken $p_r= \rho_r/3$. 

To work with the fluctuations it is more convenient to use the
following set of dimensionless quantities: 

\begin{eqnarray}
\zeta^\phi&=&H\delta\phi/\dot\phi,  
\label{zetaf}\\ 
\zeta^d&=&-\delta\dot\phi/\dot\phi, 
\label{zetad}\\ 
\zeta^r&=&\delta\rho_r/4\rho_r, \label{zetar}\\ 
\zeta^v&=&-H\delta
v^{r}  \label{zetav}\,.
\end{eqnarray}
Note that $\zeta^\phi$ and $\zeta^v$ are, respectively, the field and
radiation comoving curvature perturbation in the constant curvature
gauge, while $\zeta^r$ is the radiation curvature perturbation in the
uniform density gauge. 

The dissipation coefficient $\Upsilon$ depends on the temperature and
the scalar field
\cite{Berera:2008ar,BasteroGil:2010pb,BasteroGil:2012cm}. But at
leading order in the slow-roll approximation,  only the temperature
dependence $\Upsilon \propto T^c$ is relevant:  

\begin{equation}
\Upsilon^{-1}\delta_1\Upsilon=c\,T^{-1}\delta_1 T=c\,\zeta_1^r,
\end{equation}
where $\zeta^r=\delta\rho^N/4\rho_r$, and $\rho_r \propto T^4$. The
second-order variation makes use of the relationship (\ref{rhone}),

\begin{equation}
\Upsilon^{-1}\delta_2\Upsilon=c\,\zeta_2^r+c_2(\zeta_1^r)^2
-\frac13cH^{-2}(\partial_\alpha\zeta_1^v)(\partial^\alpha\zeta_1^v),
\end{equation}
where $c_2=c(c-4)/2$.   The  frame transformations are also used for
the second-order variation of  $D\phi=u^a\nabla_a\phi$,

\begin{equation}
\delta_2(D\phi)=\delta_2\dot\phi - H^{-1}
(\partial^\alpha\zeta^v)(\partial_\alpha\delta_1\phi)
+\frac12H^{-2}(\partial_\alpha\zeta_1^v)(\partial^\alpha\zeta_1^v)\dot\phi.
\end{equation}

The conserved total stress-energy tensor is given by that of the
scalar field and the radiation fluid $T_{ab}=
T_{ab}^{r}+T_{ab}^{(\phi)}$, and we can write

\be  \nabla^a T_{ab}^{r}= J_b = -\nabla^a T_{ab}^{(\phi)} \,,  \ee
where $J_a$ is interpreted as the flux of energy and momentum from the
scalar to the fluid system~\cite{Bastero-Gil:2014jsa}. After using
eq.~(\ref{infeq}), we find

\begin{eqnarray}
J_0&=&\Upsilon D\phi\,\dot\phi- \sqrt{2 \Upsilon
  T}\xi\dot\phi\,,\\ J_\alpha&=&\Upsilon\dot\phi\,\partial_\alpha\phi-
\sqrt{2 \Upsilon T} \xi\partial_\alpha\phi\,. 
\end{eqnarray}
The noise terms in the energy-momentum flux vector were a new feature
introduced in  ref.~\cite{Bastero-Gil:2014jsa}. We shall present
results later for two cases, one with and one without the noise terms
in the fluxes. This way we will be able to clearly see the effects
that result from the noise term.

\subsection{First-order perturbations}

If we drop the metric perturbations and the derivatives of the
potential (see the appendix~\ref{appA}), then the first-order inflaton
equation in the normal frame (\ref{infeq}) becomes 

\begin{equation}
\delta_1\ddot\phi+3H\delta_1\delta\phi+\delta_1(\bar \Upsilon D\phi)
-\partial^2\delta_1\phi=(2\bar \Upsilon T)^{1/2}\xi, 
\end{equation}
where we have replaced $\Upsilon$ by $\bar\Upsilon\equiv
\bar\Upsilon({\bf k})$, which includes the  dependence of the
dissipation coefficient on the momentum, as appropriate when treating
perturbations instead of background quantities (see the
appendix~\ref{appB} for details).  The radiation equation without the
metric perturbations is

\begin{equation}
\delta_1\dot\rho^N+4H\delta_1\rho^N+\frac43\rho_r\partial^2\delta_1v^N
=\delta_1J_0.
\end{equation}
Similarly, the scalar velocity perturbation satisfies

\begin{equation}
\frac43 a^{-3}\left\{a^3\rho_r\delta_1 v^N\right\}\dot{\phantom{A}}
+\frac13\delta_1\rho^N= -\partial^{-2}\partial^\alpha\delta_1
J_\alpha.
\end{equation}

Using the dimensionless variables introduced in
eqs.~(\ref{zetaf})-(\ref{zetav}), the first-order fluctuation
equations can be written in a compact form as: 

\begin{equation}
L(a,H,\phi,\dots)\zeta^i=K^i\xi^i,
\end{equation}
where $L=H^{-1} \partial_t +\ldots$. The slow-roll approximation
allows us to use the values of the fields at horizon crossing together
with an expansion in slow-roll parameters $\epsilon_X=d\ln X/Hdt$.
The leading-order slow-roll approximation is defined by setting
$\epsilon_X=0$, and then $L$ depends only on the background values of
the fields at horizon crossing and the value of $k$, in the
combination $z=k/(aH)$.   {}Furthermore, in constant curvature gauge,
the metric fluctuations are of order $\epsilon_H$ times $\zeta^i$ (at
first and second perturbative order) and drop out of the equations for
the fluctuations at leading-order in the slow-roll parameters
(appendix~\ref{appA}). Therefore, the dimensionless form of these
equations at leading-order, in momentum space, are

\begin{eqnarray}
H^{-1}\dot\zeta_1^\phi+\zeta_1^d&=&0,\label{pert1}\\ 
H^{-1}\dot\zeta_1^d+3(1+\bar
Q)\zeta_1^d-z^2\zeta_1^\phi-3\bar
Qc\,\zeta_1^r&=&-K^d\xi,\\ 
H^{-1}\dot\zeta_1^r-(c-4)\zeta_1^r+2\Gamma\zeta_1^d+\frac13z^2\zeta_1^v&=&K^r\xi,\\ 
H^{-1}\dot\zeta_1^v+3\zeta_1^v-\zeta_1^r-3\Gamma\zeta_1^\phi&=&0,
\label{pert4}
\end{eqnarray}
where: 

\begin{equation}
\langle\xi(k,z)\xi(k',z')\rangle=H^{-2}a^{-3}(2\pi)^3\delta(k+k')\delta(t-t').
\label{xixi}
\end{equation}
As mentioned before, the dissipation coefficient carries a momentum
dependence (appendix~\ref{appB}), $\bar\Upsilon \equiv
\bar\Upsilon({\bf k}) = \Gamma \Upsilon$, where $\Gamma \approx
e^{-k/(2 a T)}$. This explicit momentum dependence is used in  $\bar
Q=\Gamma Q$.  The dimensionless noise coefficients are
\begin{equation}
K^d={H\dot\phi\over \rho^{(\phi)}+p^{(\phi)}}(2\bar \Upsilon
T)^{1/2},\quad K^r=-{H\dot\phi\over 3(\rho_r+p_r)}(2\bar \Upsilon
T)^{1/2} \,.
\label{Kr}
\end{equation}
Note that the noise and damping terms become unimportant in the
large-scale limit $z=k/(aH)\to 0$.

The gauge-invariant variable depends on the total density perturbation,

\begin{equation}
\delta_1\rho=\delta_1\rho^N+\dot\phi\,\delta_1\dot\phi+V_\phi\delta_1\phi,
\end{equation}
to leading-order in the slow-roll parameters.  The gauge-invariant variable
can, therefore, be expressed in terms of the scalar and fluid
variations, using $p+\rho=(1+Q)\dot\phi^2$,

\begin{equation}
\Phi={Q\over 1+Q}\zeta_1^r-\frac13{1\over
  1+Q}\zeta_1^d-\zeta_1^\phi \,.
\label{zetaB1}
\end{equation}

At late times, when $z\to 0$, we have
$\zeta_1^B=-\zeta_1^\phi=-\zeta_1^v$.  However, the leading-order
slow-roll approximation is valid for $z^2>\epsilon_H$. The range of
validity overlaps the large-scale regime ($z<1$), where the gauge
invariant variables are constant. Therefore, we can solve the
leading-order slow-roll equations and match the result on the large
scale approximation in the range  $\epsilon_H^{1/2}<z<1$.

\subsection{Second-order perturbations}

Similarly, if we drop the metric perturbations and the derivatives of
the potential, then the second-order inflaton equation in the normal
frame becomes 

\begin{equation}
\delta_2\ddot\phi+3H\delta_2\delta\phi+\delta_2(\Upsilon D\phi)
-\partial^2\delta_2\phi=\delta_1 K\xi^\phi.
\end{equation}
The radiation equation is

\begin{equation}
\delta_2\dot\rho^N+4H\delta_2\rho^N+\frac43\rho_r\partial^2\delta_2v^N
=\delta_2J_0.
\end{equation}
{}Finally, the scalar velocity perturbation satisfies

\begin{equation}
\frac43 a^{-3}\left\{a^3\rho_r\delta_2  v^N\right\}\dot{\phantom{A}}
+\frac13\delta_2\rho^N-
\partial^{-2}\partial^\alpha\partial^\beta\delta_2\Pi_{\alpha\beta}=
-\partial^{-2}\partial^\alpha\delta_2 J_\alpha\;,
\end{equation}
where $\delta_2\Pi_{\alpha\beta}$ is given in eq.~(\ref{pi}).

In the slow-roll approximation, the total second-order density
perturbation is a combination of terms, 

\begin{equation}
\delta_2\rho=\delta_2\rho^N+\dot\phi\delta_2\dot\phi+V_\phi\delta_2\phi
+\frac12(\delta_1\dot\phi)^2+(\partial_\alpha\delta_1\phi)(\partial^\alpha\delta_1\phi)
\,,
\end{equation}
and the gauge invariant variable is given by

\begin{equation}
(1+Q)\Phi=Q\zeta_2^r-\frac13\zeta_2^d-(1+Q)\zeta_2^\phi
  +\frac16(\zeta_1^d)^2+\frac13H^{-2}
  (\partial_\alpha\zeta_1^\phi)(\partial^\alpha\zeta_1^\phi) \,.
\label{zetaB2}
\end{equation}

\section{Bispectrum: numerical scheme}
\label{sec3}

{}Following the prescription for the first- and second-order
perturbations discussed in the previous section, we find that the
fluctuation equations can, therefore, be realized as a linear system
of differential equations for a set of dimensionless fluctuating
quantities $\zeta^i$. If we use  $\zeta_1^i$ for the first-order
terms and $\zeta_2^i$ for the second-order, then

\begin{eqnarray}
L\zeta_1^i&=&K^i\xi^i,\label{sde}\\ L\zeta_2^i&=&j^i(k,\zeta_1,\xi),\label{de}
\end{eqnarray}
where $L=H^{-1}\partial_t+\dots$.  The source terms $j^i$ can be read
from the second-order equations given in the previous section. Sample
source terms can be expressed in the simple forms,

\begin{eqnarray}
&&c_1{}^i{}_{pq}\,\zeta_1^p\star\zeta_1^q\label{c1},\\ 
&&c_2{}^i{}_{pq}\,k^{-2}(k^\alpha\zeta_1^p\star
  k_\alpha\zeta_1^q),\\ &&c_3{}^i{}_{pq}\,k^{-2}(\zeta_1^p\star
  k^2\zeta_1^q),\\ 
&&c_4{}^i{}_{pq}\, K^i\zeta_1^p\star
  \xi^q,\\ 
&&c_5{}^i{}_{pq}\,k^{-2}k^\alpha(k_\alpha\zeta_1^p\star
  \xi^q),
\label{c5}
\end{eqnarray}
where repeated $p$ and $q$ indices are summed. These coefficients are
given explicitly in Table~\ref{table1}. 

\begin{table}[htb]
\begin{tabular}{|l|l|l|l|l|l|}
\hline \hline $c_1{}^d{}_{dr}=-3cQ$ &   $c_1{}^d{}_{rr}=3c_2Q$& 
$c_1{}^r{}_{dd}=1$ &  $c_1{}^r{}_{dr}=-2c$ &   $c_1{}^r{}_{rr}=c_2$&
 \\ &&&&&\\
$c_2{}^d{}_{\phi v}=3Q$    &   $c_2{}^d{}_{vv}=(c-3/2)Q$ & 
$c_2{}^r{}_{\phi v}=1$ & $c_2{}^r{}_{vv}=(c/3-1/2)$ & 
$c_2{}^v{}_{\phi r}=3c$ &  $c_2{}^v{}_{\phi d}=-3$ 
\\ &&&&&\\
$c_3{}^v{}_{\phi r}=3c$ &  $c_3{}^v{}_{\phi d}=-3$ & &  & &   

\\ &&&&&\\ 
$c_4{}^d{}_{r \phi }=-c/2$&   $c_4{}^r{}_{d \phi}=1$& &  & &  
\\ &&&&&\\ 
$c_5{}^r{}_{r \phi}=-c/2$ &
&   $c_5{}^v{}_{\phi\phi}=-3$ & $c_5{}^v{}_{rv}=-1/2$ & &  
\\ \hline \hline
\end{tabular}
\caption{Coefficients of the quadratic terms in the second-order
  equations. $c_2=c(c-4)/2$.}
\label{table1}
\end{table}

On large scales we use a gauge independent variable $\Phi$,

\begin{eqnarray}
\Phi_1&=&c_i\zeta_1^i,\\ \Phi_2&=&c_i\zeta_2^i+b_{ij}\zeta^i\star\zeta^j+
e_{ij}k^{-2}(k^\alpha\zeta_1^i\star k_\alpha\zeta_1^j)\,.
\end{eqnarray}
In the late time limit, it is always possible to choose $c_i$ to be
constant.  {}For our particular choice of the gauge invariant variable, 
the coefficients $c_i$, $b_{ij}$, $e_{ij}$ can be
read from eqs.~(\ref{zetaB1}) and (\ref{zetaB2}):

\bea c_r&=& \frac{Q}{1+Q} \,,\;\;\;c_d=-\frac{1/3}{1+Q}
\,,\;\;\;c_\phi=-1\,, \nonumber \\ b_{dd}&=& \frac{1/6}{1+Q} \,,
\nonumber \\ e_{\phi \phi} &=&- \frac{z^2/3}{1+Q} \,.  \eea

A simple numerical scheme for calculating the bispectrum proceeds as
follows. Choose a time $t_f$ at which all the scales of interest have
left the horizon.  The bispectrum for the gauge invariant variables
can be calculated from

\begin{equation}
B(k_1,k_2,k_3)\delta(\Sigma k)=\sum_{\rm cyc} \langle
\Phi_1(k_1,t_f)\Phi_1(k_2,t_f)\Phi_2(k_3,t_f)\rangle,
\end{equation}
where `cyc' denotes the set of cyclic permutations. We shall construct
a differential equation for the bispectrum, which relates it to two-point
correlation functions.

The first-order correlation of $\Phi_1$ with $\zeta^i_1$ is denoted by $F^i$, 

\begin{equation}
\langle\Phi_1(k_1,t_f)\zeta^j_1(k_2,t)\rangle=
k_1^{-3}(2\pi)^3\delta(k_1+k_2)F^j(k_1,t).
\label{deff}
\end{equation}
The power spectrum for $\Phi$ is related to $F^i$ by

\begin{equation}
P_\Phi(k)=k^{-3}c_iF^i(k,t_f)\,.
\end{equation}
Begin with the power spectrum calculation using eq.~(\ref{sde}). By
placing the system  in a periodic box of length $l$ it is possible to
replace $\delta(0)$ with $l^3$. Rescale the variables as follows,

\begin{equation}
\hat\zeta^i=(k/l)^{3/2}\zeta^i,\quad \hat\xi^i=(k/l)^{3/2}\xi,
\label{senorm}
\end{equation}
then $\hat\xi$ is a Gaussian random variable with

\begin{equation}
L\hat\zeta^i=K^i\hat\xi.
\end{equation}
and from eq.~(\ref{xixi}),

\begin{equation}
\langle\hat\xi(k,t)\hat\xi(k,t')\rangle=k^3a^{-3}H^{-2}\delta(t-t').
\end{equation}
The power spectrum is given by

\begin{equation}
P_\Phi(k)=k^{-3}c_ic_j\langle\hat\zeta^i(t_f)\hat\zeta^j(t_f)\rangle.
\end{equation}
We also have correlation functions for $\zeta^i$ and for the noise
$\xi^i$,

\begin{eqnarray}
F^j(k,t)&=&c_i\langle\hat\zeta^i(k,t_f)\hat\zeta^j(k,t)\rangle\label{newf},\\ 
F_\xi^j(k,t)&=&c_i
K^j \langle\hat\zeta^i(k,t_f) \hat\xi^j(k,t)\rangle.
\end{eqnarray}
Note that all the dependence on the regularization scale $l$ has
dropped off from the equations.

The full bispectrum can be found by solving an ordinary differential
equation.  The first step is to split the gauge invariant perturbation
into linear and quadratic parts,

\begin{equation}
\Phi_2=\Phi_2^{(1)}+\Phi_2^{(2)}\,.
\end{equation}
Define $B^{(1)}(k_1,k_2,k_3,t)$ as follows,

\begin{equation}
B^{(1)}(k_1,k_2,k,t)\delta(\Sigma k)= \langle
\Phi_1(k_1,t_f)\Phi_1(k_2,t_f)\Phi^{(1)}_2(k,t)\rangle \,. 
\end{equation}
{}From eq.~(\ref{de}), this satisfies

\begin{equation}
LB^{(1)}\,\delta(\Sigma k)=c_i \langle
\Phi_1(k_1,t_f)\Phi_1(k_2,t_f)j^i(k,\zeta_1)\rangle\,.
\end{equation}
The source terms are given by eqs.~(\ref{c1})-(\ref{c5}), and the
expectation values decompose into products of correlation functions,

\begin{eqnarray}
LB^{(1)}&=&
2c_ic_1{}^i{}_{pq}k_1^{-3}k_2^{-3}F^p(k_1,t)F^q(k_2,t)
\nonumber\\ 
&&+2c_ic_2{}^i{}_{pq}k_1^{-3}k_2^{-3}k^{-2}{\bf
  k_1}\cdot{\bf k_2}F^p(k_1,t)F^q(k_2,t)
\nonumber\\ 
&&+c_ic_3{}^i{}_{pq}k_1^{-3}k_2^{-3}k^{-2}\left[k_2^2F^p(k_1,t)F^q(k_2,t)+k_1^2F^p(k_2,t)F^q(k_1,t)\right]
\nonumber\\ 
&&+c_ic_4{}^i{}_{pq}k_1^{-3}k_2^{-3}\left[F^p(k_1,t)F_\xi^q(k_2,t)+F^p(k_2,t)F_\xi^q(k_1,t)\right]
\nonumber\\ 
&&+c_ic_5{}^i{}_{pq}k_1^{-3}k_2^{-3}k^{-2}{\bf k_3}\cdot
\left[{\bf k_1}F^p(k_1,t)F_\xi^q(k_2,t)+{\bf
    k_2}F^p(k_2,t)F_\xi^q(k_1,t)\right].
\label{beq}
\end{eqnarray}
The boundary conditions are $B^{(1)}=0$ at the initial time.  The
differential equation has to be solved for each set of momenta. Note
that the equation for $B^{(1)}$ has no stochastic source terms, and
all the statistical averaging is already done when constructing $F^p$.

The remaining part of the bispectrum can be obtained directly,

\begin{equation}
B^{(2)}(k_1,k_2,k)\,\delta(\Sigma k)= b_{ij}\langle
\Phi_1(k_1,t_f)\Phi_1(k_2,t_f)\zeta^i\star\zeta^j(k,t_f)\rangle +\dots
\end{equation}
After decomposing the four-point function into correlators,

\begin{equation}
B^{(2)}=2k_1^{-3}k_2^{-3} \left[ b_{ij}F^i(k_1,t_f)F^j(k_2,t_f)+
  e_{ij}k^{-2}k_1\cdot k_2 F^i(k_1,t_f)F^j(k_2,t_f)\right] \,,
\end{equation}
and the bispectrum is given by

\begin{equation}
B(k_1,k_2,k_3)=\sum_{\rm cyc}\left[
  B^{(1)}(k_1,k_2,k_3,t_f)+B^{(2)}(k_1,k_2,k_3)\right] \,.
\end{equation}

There is an important limitation of this result, which is caused by
the use of the  slow-roll approximation. We mentioned earlier that the
fluctuations stabilize and the neglected slow-roll terms are small
when $\epsilon_H<z^2<1$. This has to be true for all of the $k$ values
simultaneously, and so for example $k_1>k_2\epsilon_H^{-1/2}$.  This
cuts out the squeezed triangles with very small $k_1$ (or $k_2,k_3$).
In the squeezed triangle limit, the argument of Maldacena
\cite{Maldacena:2002vr} applies to warm inflation, and the bispectrum
must be of the order of $n_s-1$, where $n_s$ is the spectral index.
We will truncate the bispectrum for squeezed triangles.

\begin{figure}[t]
\begin{center}
\begin{tabular}{ccc}
\includegraphics[width=80mm]{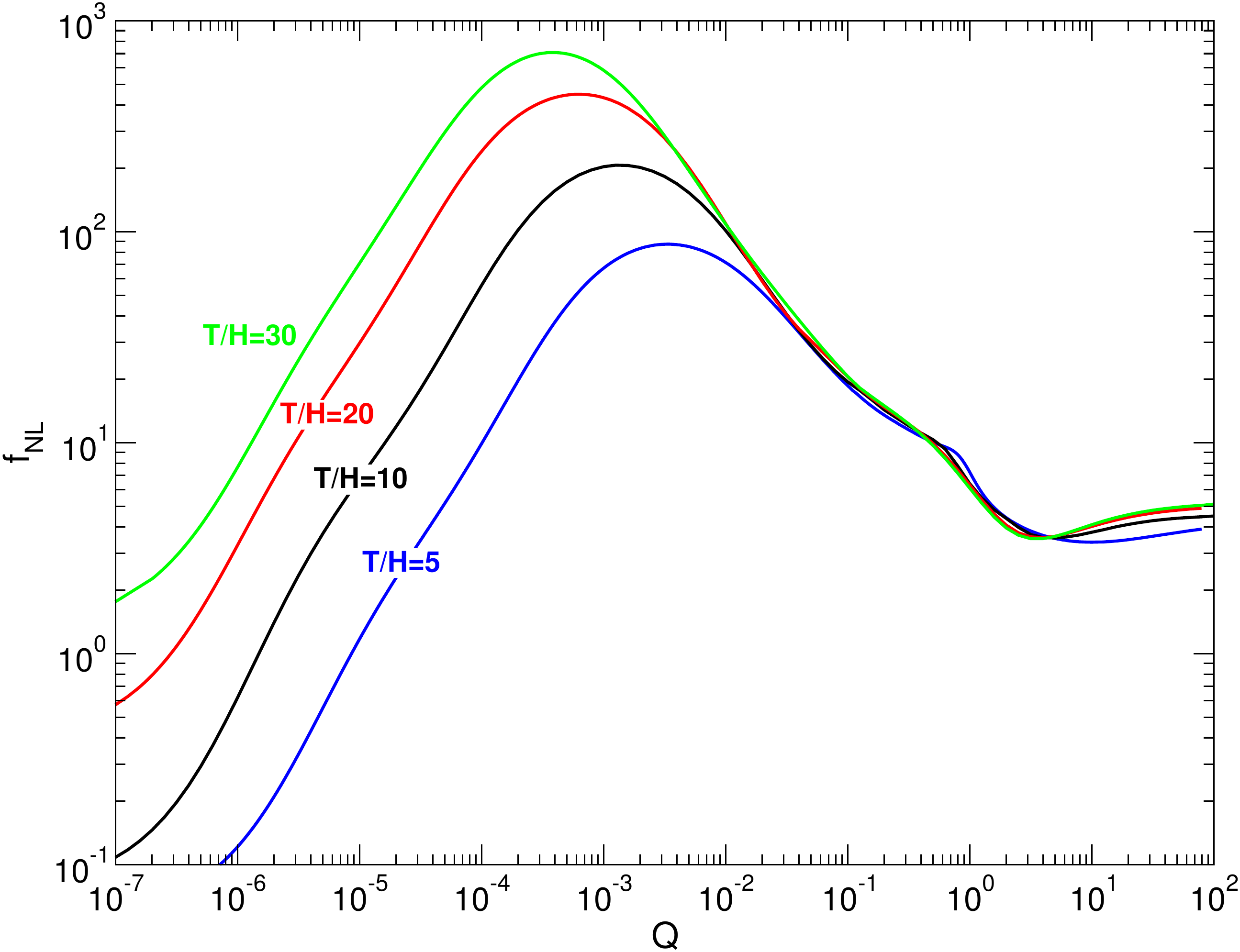} & &
\includegraphics[width=80mm]{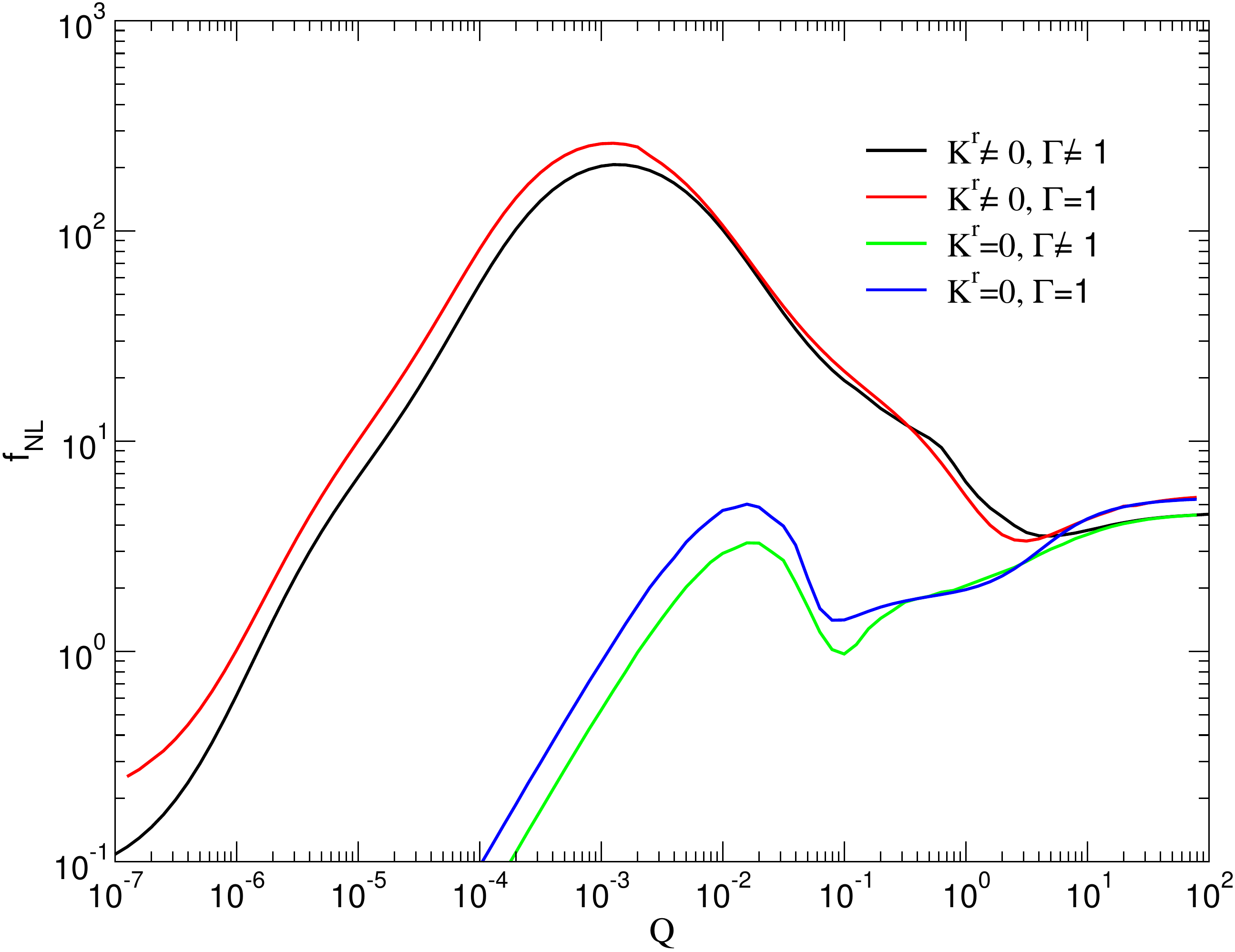} 
\end{tabular}
\end{center}
\caption{LHS: Non-linearity parameter $f_{NL}$ versus $Q$ for
  different values of $T/H$ as indicated in the plot. The dissipative
  coefficient includes a cut-off function when $z \gg T$, and the
  noise amplitude $K^r$. RHS: Comparison of $f_{NL}$ when switching
  on/off the cut-off function $\Gamma(z)$, and the noise amplitude
  $K^r$, for $T/H=10$.}
\label{plotfnl} 
\end{figure}

The magnitude and shape of the bispectrum can be reduced to a
non-linearity parameter $f_{NL}$, for which we take equilateral
triangle shapes $k_1=k_2=k_3$, and a shape function  $\overline
B(k_1,k_2,k_3)$ that factors out from the equilateral triangle
result~\cite{fNL}.  In our variables,

\be f_{NL}=\frac{18}{5}{B(k,k,k)\over P(k)^2}.  \ee Different
inflationary models predicts different shapes and
magnitudes~\cite{Fergusson:2008ra}. Observational constraints on
$f_{NL}$ depend on the shape template used. This is reflected in the
latest Planck  constraints on $f_{NL}$ \cite{planckng}, for example

\be f_{NL}^{\rm local} = 2.7 \pm 5.8 \,,\;\;\; f_{NL}^{\rm equi} = -42
\pm 75 \,,\;\;\; f_{NL}^{\rm warm} = 4 \pm 33 \,.  \ee These shapes
are described in the next section.

In the case of warm inflation, using the slow-roll approximation, the
theoretical prediction for the amplitude of the bispectrum depends only
on the dissipative ratio $Q=\Upsilon/(3H)$ and on the temperature of
the thermal bath $T/H$ at horizon crossing. In {}fig.~\ref{plotfnl}
(LHS plot) we have plotted $f_{NL}$ when varying $Q$ for different
values of $T/H$, and $c=3$, i.e., $\Upsilon \propto T^3$.

When $Q\agt 0.1$, the level of non-Gaussianity is practically
independent of $T/H$, and shows a very mild dependence on $Q$ for
$Q>1$. Strong inflationary models, which are the ones with large $Q$,
produce robust predictions for the non-Gaussianity, provided that
$c>0$. This behavior of $f_{NL}$ is due to the presence of a
``growing'' mode in the spectrum for a $T$-dependent dissipative
coefficient, which enhances the amplitude of the primordial spectrum
by a factor $Q^\alpha$, \cite{warmgrowing,BasteroGil:2011xd},  and
through the coupling of the radiation and field fluctuations enhances
the bispectrum by a factor $Q^{2\alpha}$ and, therefore, the effect
partially cancels out in $f_{NL}$. 

On the other hand, when $Q$ is small there is a strong dependence on
both $Q$ and $T/H$. We have $f_{NL} > 10$ for $10^{-3} \alt Q \alt
10^{-1}$, with the lower end for $Q$ depending on $T/H$, and being,
therefore, model dependent. Information on non-Gaussianity (combined
with that of the primordial spectrum) can then be used
in conjunction with model building to set constraints on the
inflaton interactions.

The numerical result also shows that $f_{NL}$ has a maximum at 
at around $Q\sim 10^{-3}$, approximately at the value when dissipation 
starts dominating the primordial spectrum instead of the vacuum
fluctuations \cite{Bastero-Gil:2014jsa}. For small values of $Q
\lesssim 0.1$, we have for the amplitude of the primordial spectrum: 
\be
P_\zeta \simeq \left(\frac{H}{\dot \phi} \right)^2 \left( \frac{H}{2 \pi}
\right)^2 \left( 80 \pi \frac{T}{H} Q + 1 \right) \,,
\ee
and thermal fluctuations will start dominating at around $Q\sim
(H/T)/(80 \pi)$. Similarly, the bispectrum
receives contributions from radiation fluctuations and inflaton vacuum
fluctuations. The former goes as $K^r \propto Q^{-1/2}$
(Eq. (\ref{Kr})) and thus grows towards small values of $Q$, giving a
large contribution to the non-gaussianity. However
this growth is reverted by the contribution from the vacuum
fluctuations in the denominator of $f_{NL}$, giving rise to the peak
observed in the plots. 

On the RHS in {}fig.~\ref{plotfnl} we have compared different
approximations for treating the radiation fluid perturbations,
including or not both the momentum dependence function of the
dissipation coefficient on the momentum, $\Gamma(z)$, and the noise
term $K^r$. While at large $Q$ the behavior of the fluctuations does
not depend on these terms, it is relevant at low values of $Q$, the
larger effect coming from the stochastic term in the radiation fluid
$K^r$.  This is useful to show the strong dependence of the
non-Gaussianity on the microphysics of warm inflation in the
intermediate and weak dissipation regime of warm inflation, $Q \ll 1$.
We notice that when $K^r=0$, the primordial spectrum for low $Q$ is
given by \cite{Bastero-Gil:2014jsa}:
\be
P_\zeta \simeq \left(\frac{H}{\dot \phi} \right)^2 \left( \frac{H}{2 \pi}
\right)^2 \left( 2 \pi \frac{T}{H} Q + 1 \right) \,,
\ee
dissipation takes over vacuum fluctuations at slightly larger values
of $Q$, and therefore the peak in $f_{NL}$ is shifted towards the
right (with a smaller value).

\begin{figure}[ht]
\begin{center}
\includegraphics[width=80mm]{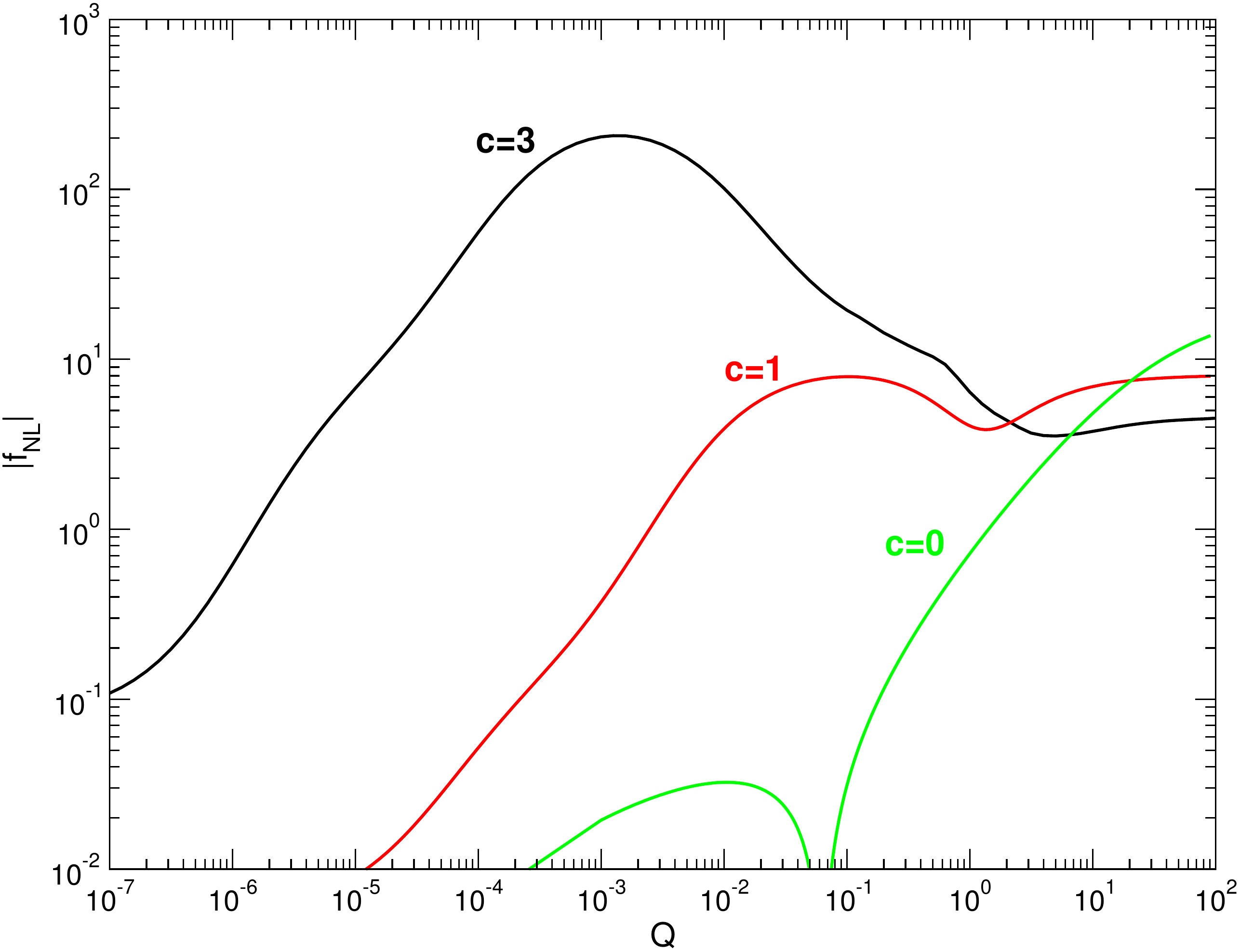} 
\end{center}
\caption{Non-linearity parameter $|f_{NL}|$ versus $Q$ for
  different values of $c$ ($\Upsilon \propto T^c$) as indicated in the
  plot. The dissipative coefficient includes wavenumber dependent function 
$\Gamma$ and the noise amplitude $K^r$  (both explained in the text). }
\label{plotfnlc} 
\end{figure}

In {}fig.~\ref{plotfnlc} we compare the value of $|f_{NL}|$ for
different values of $c$ ($\Upsilon \propto T^c$). The larger is $c$,
the larger is the coupling between radiation and field
fluctuations~\cite{warmgrowing}, which enhances the non-Gaussianity
for $Q \alt 1$. 


\section{Fitting Bispectral Shapes}
\label{sec4}

\begin{figure}[ht]
\begin{center}
\subfigure[Bispectrum for small $Q$, ($Q=10^{-4}$).]{
  \includegraphics[width=8cm]{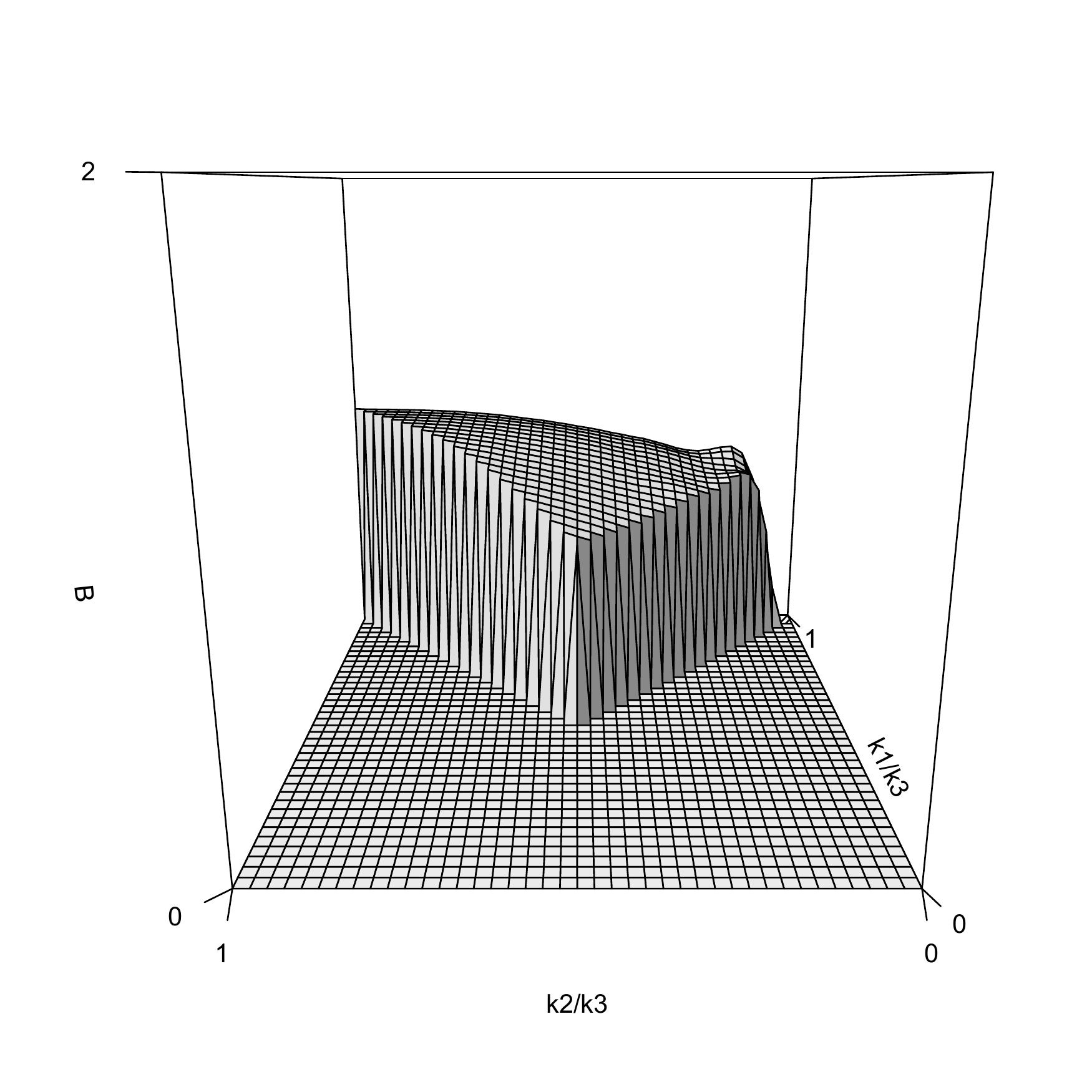}} \subfigure[Bispectrum for
  large $Q$, ($Q=100$).]{ \includegraphics[width=8cm]{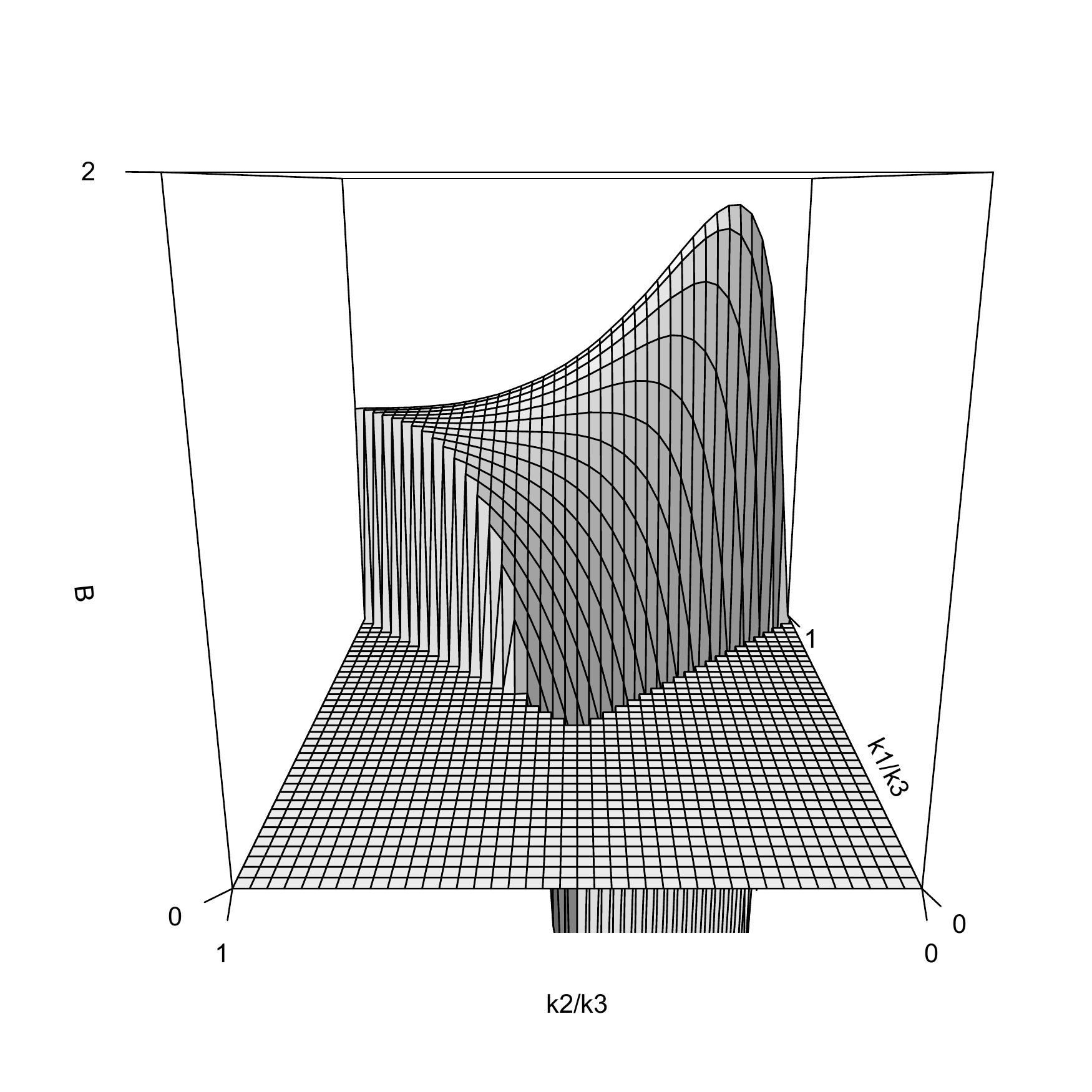}}
\end{center}
\caption{The numerical spectrum plotted against $x_1=k_1/k_3$ and
  $x_2=k_2/k_3$.}
\label{numerical} 
\end{figure}

The functional dependence of the bispectrum on the three momenta ${\bf
  k}_1$, ${\bf k}_2$ and  ${\bf k}_3$ is an important feature that can
potentially distinguish different sources of non-Gaussianity and probe
differences in inflationary models. Because of the condition ${\bf
  k}_1+{\bf k}_2+{\bf k}_3=0$, and the symmetry under permutations,
the shape can be parameterized  by parameters $x_1=k_1/k_3$ and
$x_2=k_2/k_3$, and $k_3$ can be chosen to be the  largest of the three
wave-numbers. The triangle equality implies that $x_1+x_2\ge 1$.  The
plots in figure~\ref{numerical} show the numerical results for two
values of the dissipation coefficient~$Q$.

Reconstruction of the bispectral shape from CMB observations is a
difficult  task and at present the best we might hope to do is compare
different model bispectrum templates.  We, therefore, require a
template, or a set of templates, which are indicative of warm
inflation models. Two such templates were identified in
refs.~\cite{Moss:2007cv, Moss:2007qd, Moss:2011qc} for the strong
regime of warm inflation. One of these was the same local form \cite{Salopek:1990jq} that is
obtained from other inflationary models, such as curvaton models \cite{Lyth:2005fi,Bartolo:2003bz},

\begin{equation}
B_L=\sum_{cyc}k_1^{-3}k_2^{-3} \,. 
\end{equation}
The other had the form

\begin{equation}
B_S=\sum_{cyc}k_1^{-3}k_2^{-3}(k_1^{-2}+k_2^{-2}){\bf k_1}\cdot{\bf
  k_2},
\end{equation}
which is specific to warm inflation. Note that the analytic treatment
used in refs.~\cite{Moss:2007cv,Moss:2007qd,Moss:2011qc} breaks down
for squeezed triangles, and in practice we use a truncated form of
$B_S$ which is zero if any $k_i<k_j\delta$, with $\delta\approx 0.1$
(of order of the slow-roll parameter).

\begin{figure}[ht]
\begin{center}
\subfigure[Local Bispectrum.]{ \includegraphics[width=8cm]{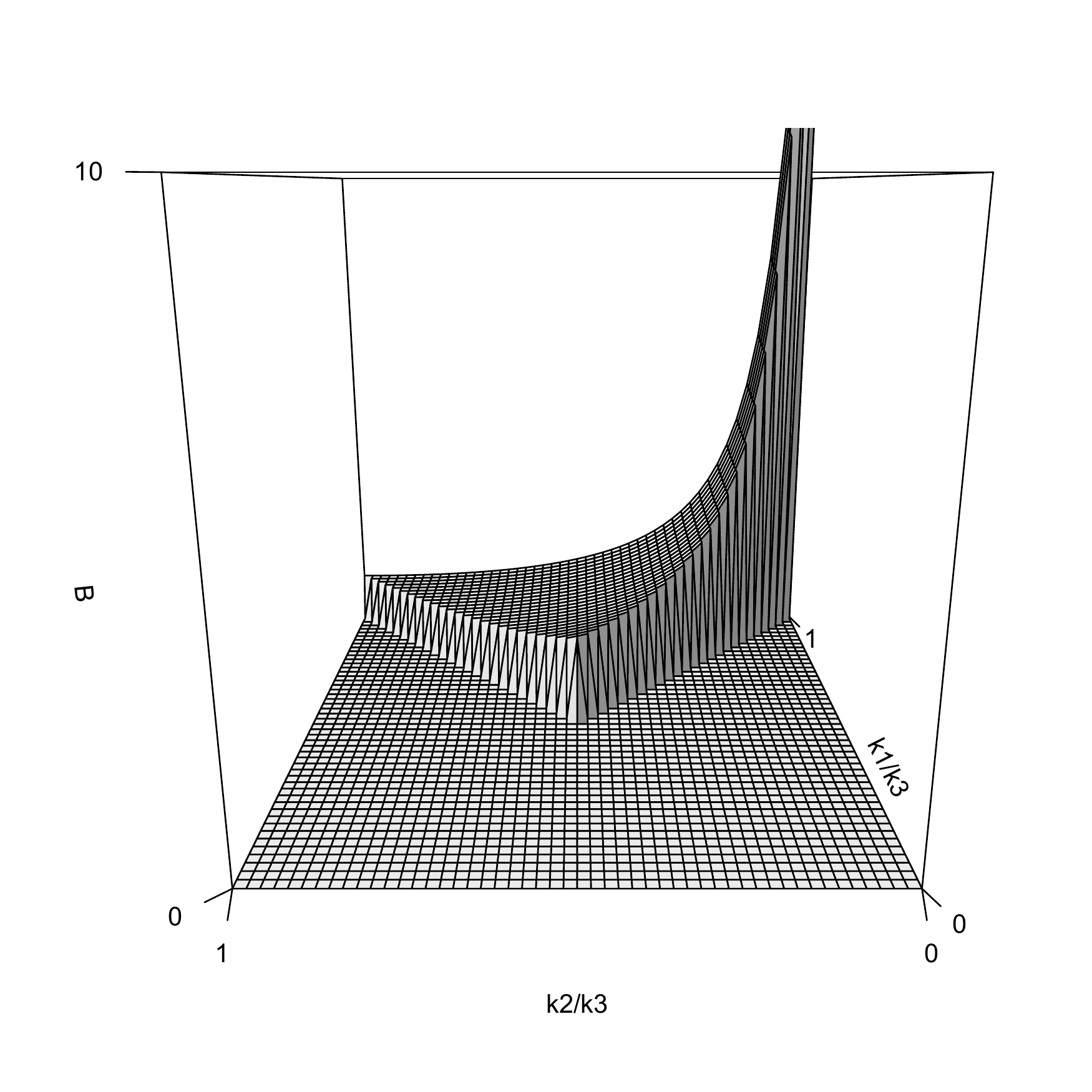}}
\subfigure[Warm bispectrum type $B_W$.]{
  \includegraphics[width=9cm]{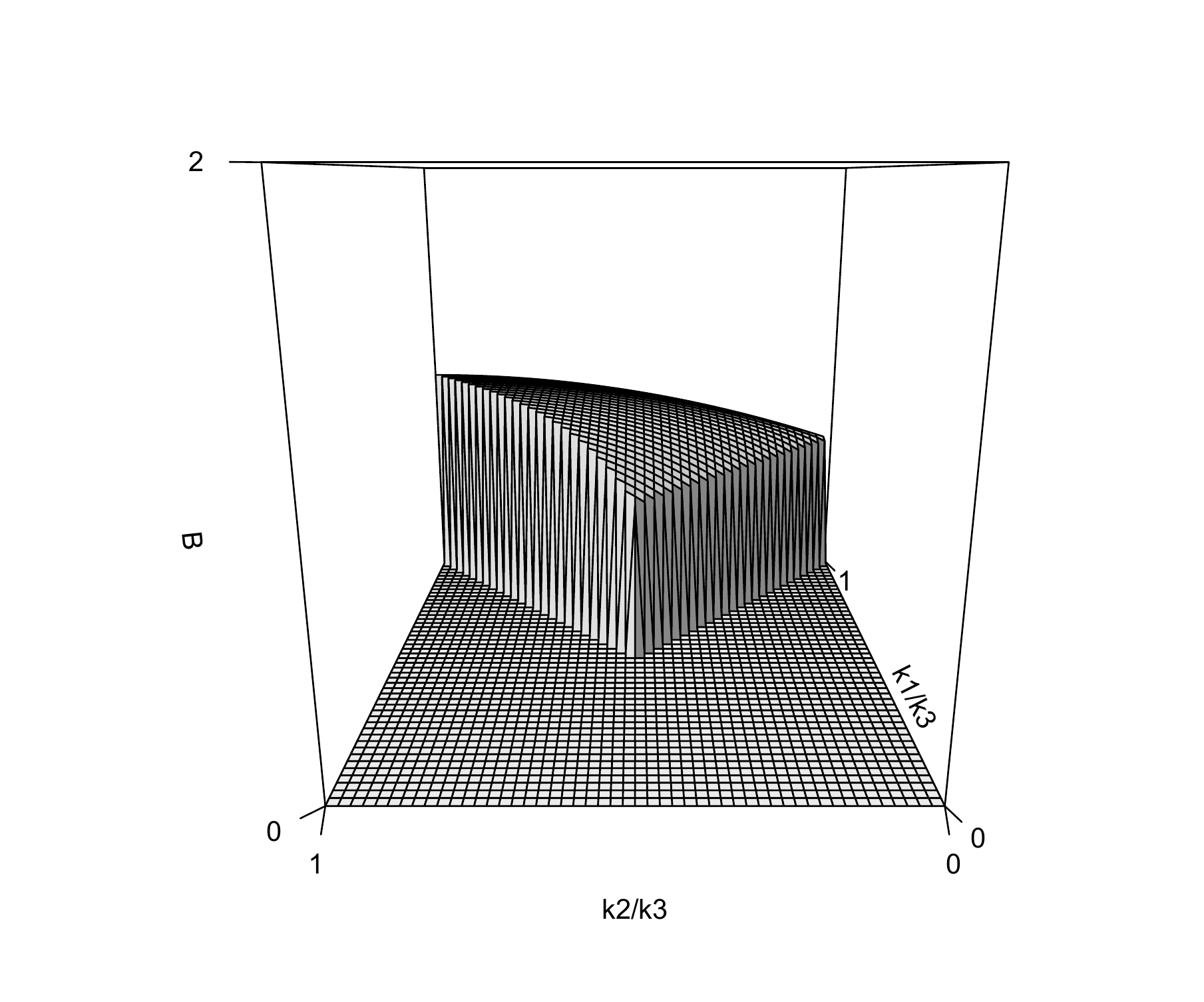}} \subfigure[Warm bispectrum type
  $B_S$.]{ \includegraphics[width=8cm]{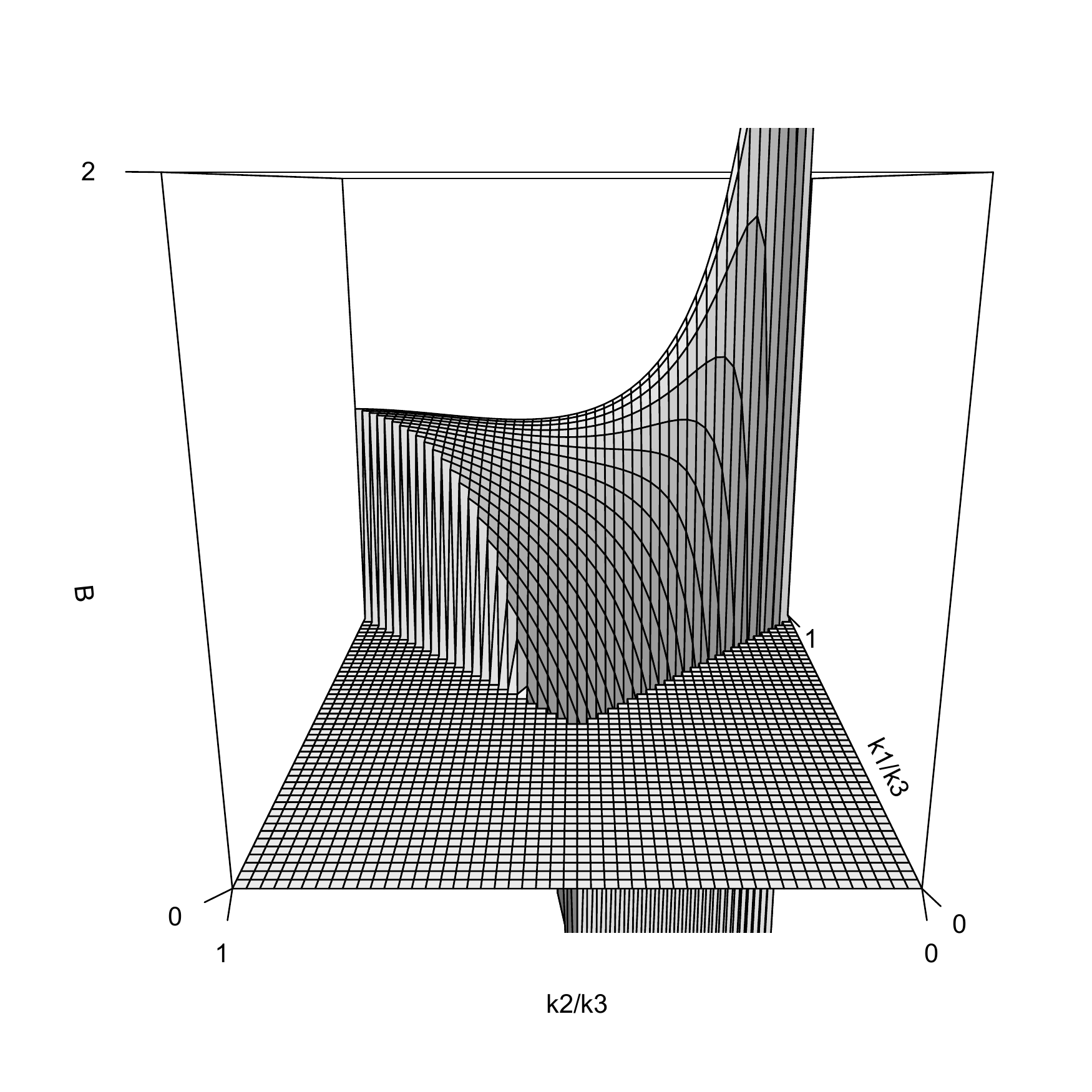}} \subfigure[Equilateral
  bispectrum.]{ \includegraphics[width=9cm]{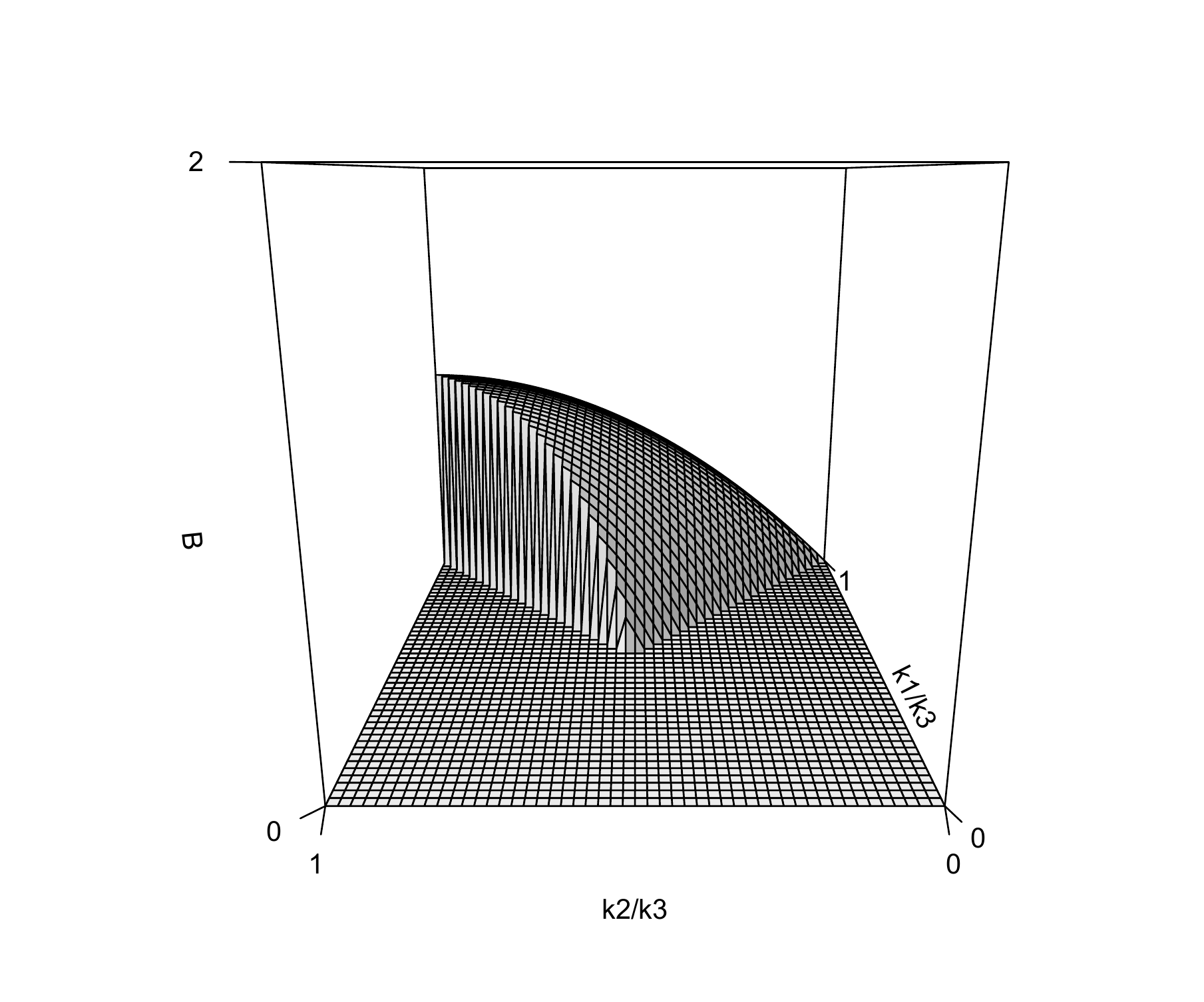}}
\end{center}
\caption{Selected bispectral shapes plotted against $x_1=k_1/k_3$ and
  $x_2=k_2/k_3$.}
\label{shapes} 
\end{figure}

Other bispectral shapes are suggested by the bispectrum equation
(\ref{beq}). If $F^p$ was constant, then the second source term would
have the form 

\begin{equation}
B_W=\sum_{cyc}k_1^{-3}k_2^{-3}k_3^{-2}{\bf k_1}\cdot{\bf k_2}\,.
\end{equation}
The first source term in eq.~(\ref{beq}) has a local shape and the
third term is a combination of the previous two when we use ${\bf
  k}_1+{\bf k}_2+{\bf k}_3=0$. These spectral shapes are plotted in
figure \ref{shapes}. The fourth shape  plotted in figure \ref{shapes}
is the equilateral template. This is similar to the warm inflation
shape $B_W$, but an important difference is that the equilateral
template vanishes when $x_1+x_2=0$, unlike $B_W$. 

{}For the weak regime of warm inflation ($Q\ll 1$), we would like to
find a simple representation of the numerical results obtained
earlier, and find a template that is an optimal fit, in some sense,
for different parameter ranges in $Q$, etc. Ideally, the comparison
between different types of bispectral function should be done on a
spherical projection using the angular components $B_{l_1l_2l_3}$
(see, e.g., ref.~\cite{Fergusson:2008ra}).  This depends on the linear
transfer function and is computationally expensive. A simpler approach
is to use a momentum space comparison as, e.g., used in
ref.~\cite{Babich:2004gb}.  Two spectral shapes with identical
momentum dependence will give the same angular components, but the
reverse is not necessarily true. The momentum space approach can be
improved by modifications of the momentum space covariance function.

Matching the numerical bispectrum to a given template requires a
distance function in bispectrum space, or equivalently, it requires an
inner product or covariance function.  This should respect the
constraints and symmetries of the bispectrum. We start from an
integral expression,

\begin{equation}
B_1\cdot B_2=\int d{\bf k}_1 d{\bf k}_2 d{\bf k}_3
{B_1(k_1,k_2,k_3)B_2(k_1,k_2,k_3)\over P(k_1)P(k_2)P(k_3)} \delta({\bf
  k}_1+{\bf k}_2+{\bf k}_3)\,.
\label{inner}
\end{equation}
It is also possible to add any function $\omega(k_1,k_2,k_3)$ that
respects the symmetries of the integrand. The integral in
eq.~(\ref{inner}) then reduces to 

\begin{equation}
B_1\cdot B_2={1\over 8\pi^4}\int_{\Delta} dk_1 dk_2 dk_3\,k_1k_2k_3\,
{B_1(k_1,k_2,k_3)B_2(k_1,k_2,k_3)\over P(k_1)P(k_2)P(k_3)}\,,
\end{equation}
where $\Delta$ is the range of the integral, restricted by the
triangle inequality. At leading order in slow-roll, the spectra have
approximate scaling symmetries of the form

\begin{equation}
B_1(k_1,k_2,k_3)=k_3^{-6}B(x_1,x_2,1),\qquad P(k)=k^{-3}P(1),
\end{equation}
where  $x_1=k_1/k_3$ and $x_2=k_2/k_3$. In the new variables, after
dropping an overall constant factor and re-instating the weight
function, the integral becomes 

\begin{equation}
B_1\cdot B_2=\int_{1/2}^1\, dx_1\int_{1-x_2}^{x_2} dx_2\,x_1^4 x_2^4\,
B_1(x_1,x_2,1)B_2(x_1,x_2,1)\omega(x_1,x_2)\,.
\end{equation}
The simplest choice $\omega(x_1,x_2)=1$ does not give convergence for
some important bispectral shapes, and so we shall choose a simple
truncation with cutoff $\delta$,

\bea 
\omega(x_1,x_2)&=& 0 \, \;\; {\rm for}\;  x_1<\delta \;\text{or}
\; x_2<\delta, 
\\ \omega(x_1,x_2)&=& 1 \,, \;\;\text{otherwise}.  
\eea
The suppression of the bispectrum for squeezed triangles is expected
in  models where the bispectrum is generated on sub-horizon 
scales~\cite{Maldacena:2002vr}.  The correlation function, or 'cosine', is
defined as the normalized product,

\begin{equation}
{\rm cor}(B_1,B_2)=\widehat B_1\cdot\widehat B_2,\qquad \widehat
B={B\over \sqrt{B\cdot B}}\,.
\end{equation}

\begin{figure}[ht]
\begin{center}
\subfigure[The correlation between the numerical bispectrum and the
  template shapes plotted as a function of the dissipation strength
  parameter $Q$.]{ \includegraphics[width=0.4\textwidth]{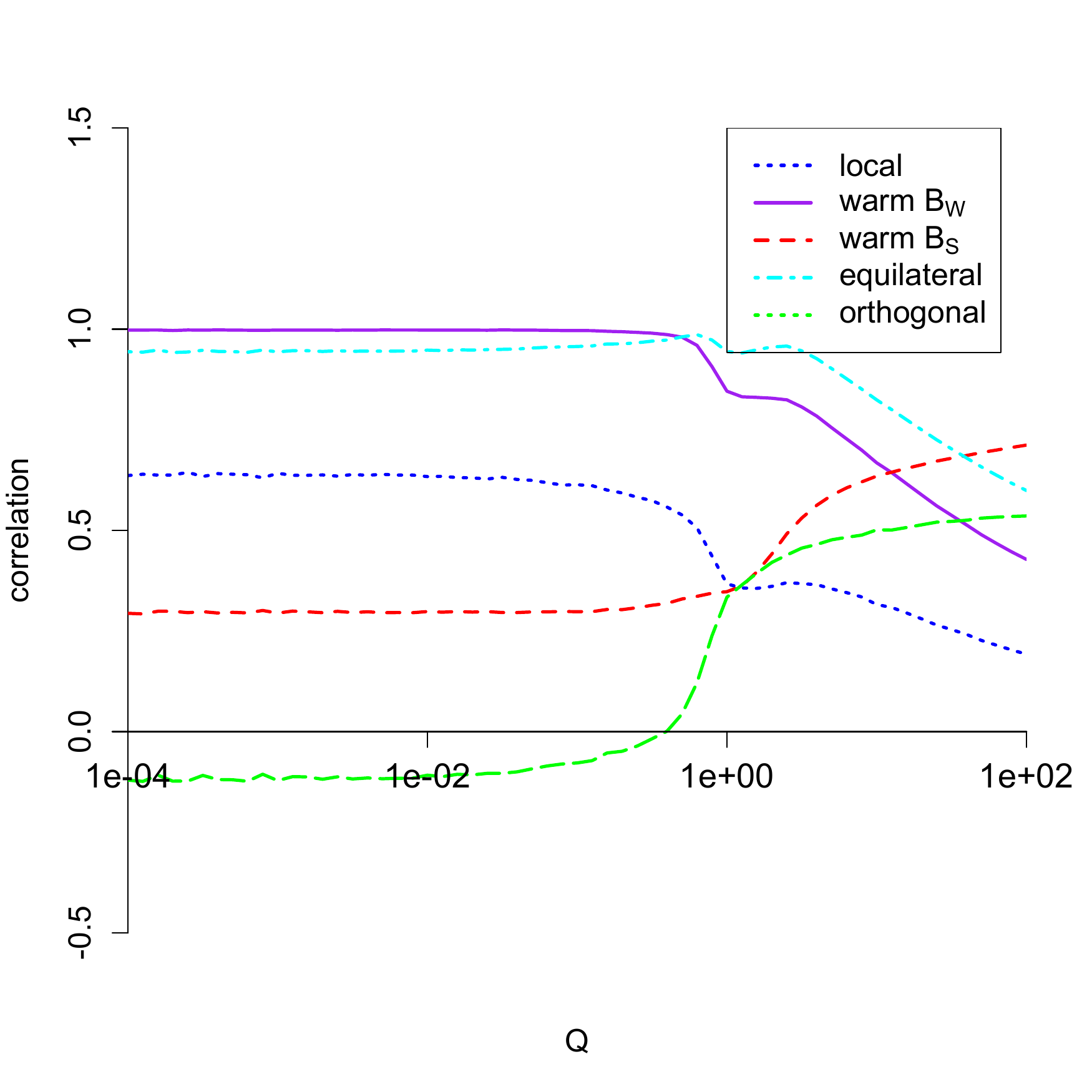}}
\subfigure[The optimal fit parameters for $B_L$, $B_W$ and $B_S$
  plotted as a function of the dissipation strength parameter $Q$.]{
  \includegraphics[width=0.4\textwidth]{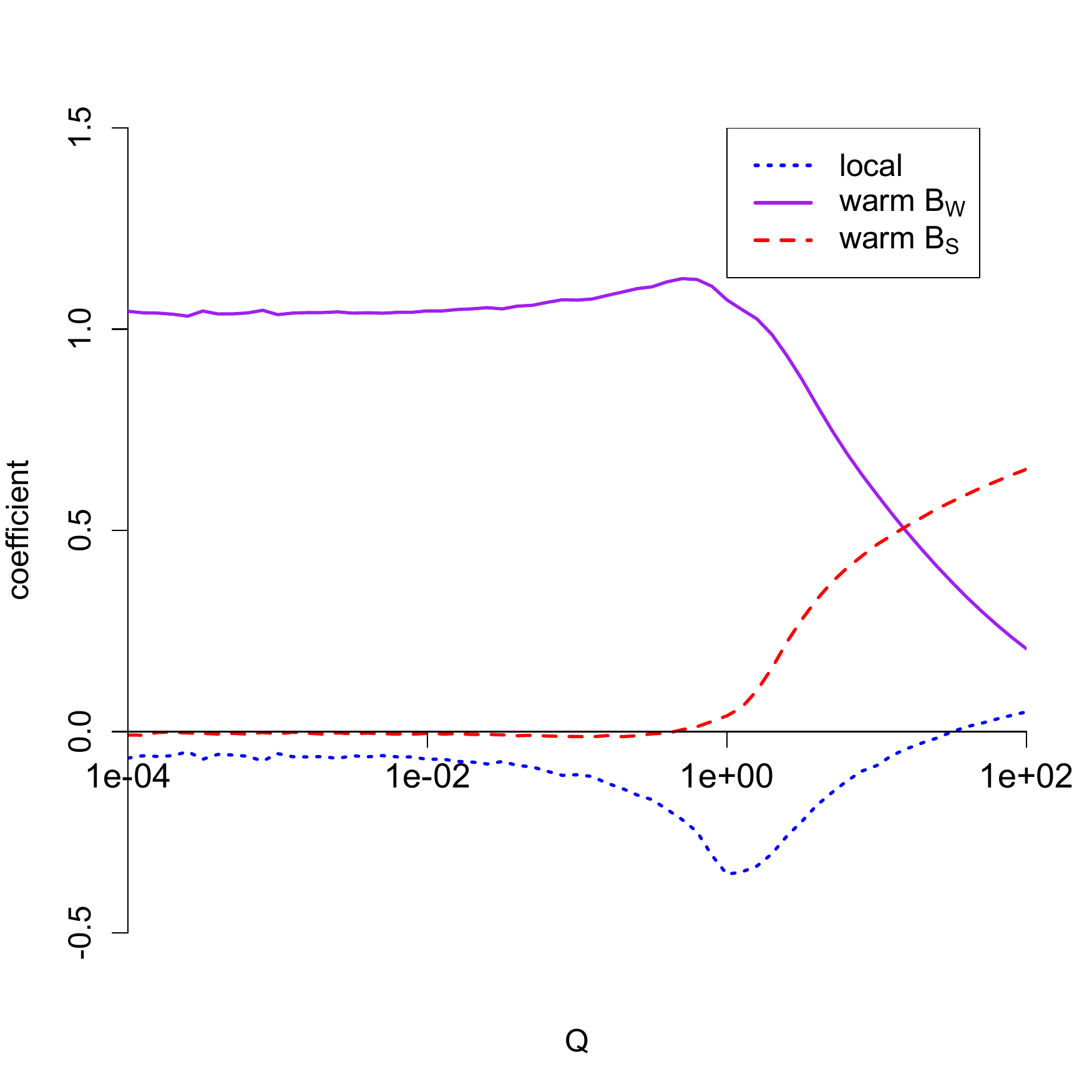}}
\end{center}
\caption{Correlation between shapes as a function of the dissipation
  strength parameter $Q$.}
\label{cor} 
\end{figure}

The correlation function has been used in figure~\ref{cor} (left-hand side panel)
to compare the numerical bispectrum for the two warm templates
$B_S$ and $B_W$ with those of the most common templates
used in the literature, namely the local template $B_L$, the equilateral
and the orthogonal shapes.
There is a clear
transition from the warm template $B_W$ in the weak regime of warm
inflation, where $Q$ is small, to the warm template $B_S$ in the
strong regime of warm inflation, where $Q$ is large. The correlation
between the equilateral and $B_W$ templates is quite large,
approximately $0.94$, and momentum space correlator cannot separate
these two shapes efficiently.

The bispectrum can also be matched to a set of templates $B_n$ with
coefficients that minimize the residuals,

\begin{equation}
E(f_n)=\left(\widehat B-\sum_n f_n \widehat B_n\right)^2\,.
\end{equation}
The square is taken using the inner product. The optimal fit has

\begin{equation}
f_n=\sum_m F^{-1}_{nm} \widehat B_m\cdot \widehat B\,,
\end{equation}
where $F^{-1}_{nm}$ is the inverse correlation matrix,
$F_{nm}=\widehat B_n\cdot \widehat B_m$. A fit to the set $B_L$, $B_W$
and $B_S$ is shown in figure~\ref{cor} (right-hand side panel).  The
equilateral template has been left out because it has a large overlap
with the $B_W$ template. The numerical bispectrum is predominantly  of
the warm $B_W$ form for small $Q$ and of the $B_S$ form for large~$Q$. 


\section{Conclusions}
\label{sec5}

If primordial non-Gaussianity is observed, we will have a powerful new
tool for distinguishing amongst the many different types of
inflationary models. The warm inflationary models form a subclass of
all inflationary models and can produce a significant amount of
non-Gaussianity in some parameter regimes.  In the {\it strong} regime
of warm inflation, the prediction for the non-linearity parameter is
$f_{NL}\approx 10$, for models with a temperature dependent
dissipation term. This is  consistent, thought slightly smaller, than
the result predicted in ref.~\cite{Moss:2011qc}, which  used crude
analytic approximations.

In the {\it intermediate} regime of warm inflation, where $Q\sim 1$, the
non-Gaussianity grows, but there is a proviso that the result depends
on the effect of the stochastic dynamics on the heat flux. The
$f_{NL}$ parameter falls off for small values of the dissipation
parameter $Q$, i.e., in the {\it weak} dissipation regime of warm
inflation, and its amplitude also depends on the temperature of the
thermal radiation bath. One should note that both the dissipation, 
temperature and the state for the inflaton fluctuations (thermal or quantum), 
strongly depend on the details of the interactions involved (see, e.g.,
refs.~\cite{Ramos:2013nsa,Bartrum:2013fia}). Thus, the amplitude of
non-Gaussianity for {\it weak} inflationary models is model dependent
and is strongly dependent on the microscopic physics and dynamics.

The most important results we have found concern the shape of the
bispectrum.  The magnitude of the non-Gaussianity is an important
observable, but the shape of the bispectrum has the potential of being
even more relevant. The more common shapes, like the equilateral,
local, flat, feature, etc., are useful to distinguish various different
models of inflation.  Nevertheless, many classes of inflation models can
be described interchangeably by some of these
shapes~\cite{planckng,Fergusson:2008ra}.  Thus, even if one of these
shapes turn out to be measured, there are still degeneracies among
these inflation models that makes distinguishing models a difficult task.
In this respect, warm inflation has specific shapes
of its own, types $B_W$ or $B_S$ as we have shown.

The shape is different for the strong and weak regimes of warm
inflation, but in both cases the shape is different from the shape of
bispectrum obtained from any other inflationary model. The weak shape
$B_W$ is quite close to the equilateral shape, and so limits on $f_{NL}$ for
the equilateral shape are most likely relevant for this regime. The
bispectral shape for the strong warm regime $B_S$ 
agrees with previous analytic
results, and this shape has a low correlation with other shapes.
Our results show that there is a clear transition from
the warm template $B_W$, which is the dominant shape in the weak
regime of warm inflation ($Q\lesssim 1$),   to the warm template $B_S$
in the strong regime of warm inflation ($Q \gtrsim 1$). This is a
novel result that has not been described in previous works.

There are some additional physical effects that can be included to
refine the analysis. One of these is the possibility of viscosity
(bulk and shear viscosities) in the radiation fluid  (for studies of
these effects  in the perturbations at the first-order see, e.g.,
refs.~\cite{Bastero-Gil:2014jsa,BasteroGil:2011xd}).  Viscosities
would affect the amplitude of power spectrum, and could also have some
effect on the bispectrum.  Though fully including these effects in the
second-order perturbation equations can be done along the lines of the
study carried out in this work, it complicates considerably the
analysis and we leave such study for a future work.
Another important consideration is that the inflaton might be in
a thermal state, adding an additional level of fluctuations beyond
those induced from the thermal radiation. These fluctuations have
been included in \cite{Bastero-Gil:2014jsa}, for example. The inclusion
of non-gaussianities arising from thermal field theory in such a 
situation is left for future work.

The results we have obtained in this paper may also be of relevance in
contexts other than warm inflation. They can be of importance, for
instance, in the studies of non-Gaussianities in curvaton type of
models and where the curvature perturbations are generated during
reheating after inflation~\cite{Ichikawa:2008ne,Leung:2012ve}. In
these cases, both dissipation and stochastic noises in the radiation
bath should be accounted for and they can be important in regards to
the magnitude of $f_{NL}$. Recall, in particular, that from the
results we have obtained here that radiation noise tends to enhance
$f_{NL}$.  This may potentially put additional pressure on
curvaton type of models, which already tend to be  in disagreement
with the recent observational results~\cite{Lyth:2014yya}.     

\acknowledgments

M.B.G. is partially supported by ``Junta de Andalucia'' (FQM101).
A.B. is partially supported by a UK Science and Technology Facilities
Council Consolidated Grant. I.G.M. is partially supported by  the UK
Science and Technology Facilities Council Consolidated  Grant
ST/J000426/1. R.O.R. is partially supported by research grants from
the  brazilian agencies  Conselho Nacional de Desenvolvimento
Cient\'{\i}fico e Tecnol\'ogico (CNPq) and  Funda\c{c}\~ao Carlos
Chagas Filho de Amparo \`a Pesquisa do Estado do Rio de Janeiro
(FAPERJ). We would like to thank the Higgs Center in Edinburgh (UK), and 
I.M. and M.B.G.the ``Centro de Ciencias Benasque Pedro Pascual'' (Spain), 
for its hospitality during the writing of this paper.  


\appendix 

\section{Metric perturbations}
\label{appA}

The spacetime metric for a scalar-type of perturbation is given by

\begin{equation}
ds^2=-(1+2\alpha)dt^2-2\beta_{,i}dt \,
dx^i+a^2\left(\delta_{ij}(1+2\varphi)+2\gamma_{,ij}\right)
dx^idx^j\,.\label{metric}
\end{equation}
The perturbed Einstein equations up to second-order can be found in
the literature (see, e.g.,
refs.~\cite{Acquaviva:2002ud,Bartolo:2003bz,Noh:2004bc,Hwang:2007ni})
and are given below. These imply that the  first-order metric
perturbations are first-order in the slow-roll parameter and the
second-order metric perturbations are second-order in the slow-roll
parameters. We shall give the results for $\alpha$ as an example.

We make use of the shear $\chi$ and perturbed expansion rate $\kappa$,
define, respectively, by

\begin{eqnarray}
\chi&=&a(\beta+a\dot\gamma),\\ \kappa&=&3H\alpha-3\dot\varphi-\partial^2\chi.
\label{defkappa}
\end{eqnarray}
The first-order perturbations of the Einstein equations are
then~\cite{Hwang:2001fb}

\begin{eqnarray}
\partial^2\varphi_1+H\kappa_1&=&-4\pi
G\delta_1\rho,\label{e1}\\  
\kappa_1+\partial^2\chi_1&=&-12\pi
G(\rho+p)\delta_1
v,\label{e2}\\  
\dot\chi_1+H\chi_1-\alpha_1-\varphi_1&=&8\pi
G\delta_1\Pi,\label{e3}\\  
\dot\kappa_1+2H\kappa_1+\partial^2\alpha_1-3(\rho+p)\alpha_1&=&
4\pi G(\delta_1\rho+3\delta_1 p).\label{e4}
\end{eqnarray}

The density, pressure and shear perturbations are the sum of the fluid
and scalar density and pressure perturbations.

In constant curvature gauge, $\varphi_1=\varphi_2=0$, these combine to
give

\begin{eqnarray}
\alpha_1&=&\epsilon_HH\delta_1 v,\\ 
{\kappa_1\over
  3H}&=&\epsilon_H{\delta_1\rho\over
  3(p+\rho)},\\ 
\chi_1&=&\partial^{-2}(3H\alpha_1-\kappa_1), 
\end{eqnarray}
where $\epsilon_H= - \dot H/H^2= 4 \pi G (\rho +p)/H^2$.  Note that
the right-hand sides of the first two equations above are the product
of a slow-roll parameter  with the Lukas and Curvature variables
respectively. The corresponding metric perturbations are explicitly
first-order in the slow-roll expansion. On large scales, for {}Fourier
modes with $k<aH$, the shear contains a growing factor $k^{-2}$ but
remains first-order as a result of the Lukas and Curvature variables
converging to the same constant value. 

The second-order equations are much more complicated, and Noh and
Hwang~\cite{Noh:2004bc,Hwang:2007ni} (in their `spatial $C=0$ gauge' with
$\gamma=0$ and $\beta_{,\alpha}= \partial_\alpha\chi$)   give an
equation for $\alpha_2$, which is

\begin{equation}
\alpha_2=\epsilon_H H\delta_2v+\frac13 H^{-1}(N_2-N_0)\,,
\end{equation}
where

\begin{equation}
N_0=-\frac92H\alpha_1^2+\alpha_1\partial^2\chi_1 +\frac32
H(\partial_\alpha\chi_1)(\partial^\alpha\chi_1)\,,
\end{equation}
and

\begin{eqnarray}
N_2&=&-\partial^{-2}\partial^\alpha(\alpha_1\partial_\alpha\kappa_1)
-3\epsilon_H\partial^{-2}\partial^\alpha(\alpha_1\partial_\alpha
H\delta_1v) 
\nonumber\\ 
&&+\frac32\partial^{-2}\partial^\alpha\left(
(\partial_\beta\alpha_1)(\partial_\alpha\partial^\beta\partial^{-2}(3H\alpha_1-\kappa_1))\right) 
-\frac12\partial^{-2}\partial^\alpha(\partial_\alpha \alpha_1 ( 3 H \alpha_1 - \kappa_1))
\,.
\end{eqnarray}
It follows that $\alpha_2$ is first-order in the slow-roll expansion,
whilst $N_2$ and $N_0$ are second-order.

\section{Dissipation coefficient ${\bar \Upsilon}$}
\label{appB}

The dissipation coefficient $\Upsilon$ in warm inflation describes the
way the inflaton transfers its energy to radiation degrees of
freedom. Its explicit form depends on the details of the interactions
involved. These include the direct coupling of the inflaton to other
fields, but also of these with other  degrees of freedom, which make
the radiation bath. Details of successful  interaction schemes were
first reported in ref.~\cite{BR1} (for details of the quantum field
theory derivation of these dissipation terms, see, e.g.,
refs.~\cite{Berera:2008ar,BasteroGil:2010pb,BasteroGil:2012cm}).
{}For example, for a typical coupling of the inflaton field $\phi$ to
other scalar fields $\chi$ of the form $g_\chi \phi \chi^2$, we can
define a  nonlocal in both space and time dissipation term entering in
the effective equation of motion for the inflaton as~\cite{BMRnoise}

\begin{eqnarray}
{\bar \Upsilon}({\bf k},\omega) = \frac{ g_\chi^2}{
  \omega \, n(\omega)} \int \frac{d^3 p}{(2 \pi)^3}
\int_{-\infty}^\infty {d\omega' \over 2\pi} n(\omega')
n(\omega-\omega') \tilde{\rho}_\chi({\bf p},\omega')
\tilde{\rho}_\chi({\bf k}-{\bf p},\omega-\omega')\;,
\label{noise1kernel}
\end{eqnarray}
where $n(\omega)$ is the Bose-Einstein distribution function and
$\tilde{\rho}_\chi({\bf k},\omega)$ is the spectral function for the
$\chi$ field,

\begin{equation}
\tilde{\rho}_\chi({\bf k}, \omega) = \frac{4 \omega_\chi({\bf k})
  \Gamma_\chi({\bf k},\omega)}{ \left[ \omega^2 - \omega_\chi^2({\bf
      k}) \right]^2 + \left[ 2  \omega_\chi({\bf k})  \Gamma_\chi({\bf
      k},\omega) \right]^2 }\;,
\label{rhotau}
\end{equation}
where $\omega_\chi({\bf k})$ is the dispersion relation for the field
$\chi$ and $\Gamma_\chi({\bf k},\omega)$ is its decay width. The
explicit expression  for $\Gamma_\chi({\bf k},\omega)$ can be found,
e.g., in ref.~\cite{BasteroGil:2010pb} for different couplings of
$\chi$ with (light) radiation fields.  

The local approximation for the dissipation coefficient, as
appropriate to describe the background evolution, is defined by taking
the limit $\omega\to 0,\; {\bf k} \to 0$  in
eq.~(\ref{noise1kernel}). In this local approximation,  ${\bar
  \Upsilon}({\bf k},\omega) \to {\bar \Upsilon}(0,0) \equiv \Upsilon$.
The perturbation for the inflaton field, however, involves an explicit
dependence on the (space) momentum. Thus, we cannot   take the local
limit in space, ${\bf k} \to 0$, in eq.~(\ref{noise1kernel}). 

{}First-order perturbations are not so much sensitive to small scales
(large wavenumbers), since it is mostly determined by those momentum
modes corresponding to scales  larger than the horizon ($z \ll
1$). This justifies the use of a local approximation for the
dissipation coefficient in previous works (e.g., in
refs.~\cite{warmgrowing,Bastero-Gil:2014jsa}). However, the (space)
momentum dependence is particularly important in the evaluation of the
non-Gaussianity, since it is most sensitive to the small scale physics
(large momentum, or $z \gtrsim 1$).  Taking the time localization of
the dissipation coefficient in eq.~(\ref{noise1kernel}), but keeping
the space momentum contribution, the most important contribution when
${\bf k} \neq 0$ comes from the decay width in
eq.~(\ref{rhotau}). With the explicit expressions found, e.g., in
refs.~\cite{BasteroGil:2010pb,BMRnoise},  for light radiation fields
coupled to the $\chi$ field, we obtain that

\begin{equation}
{\bar \Upsilon}({\bf k},\omega \to 0) \approx e^{-k/(2 a T)}
\Upsilon\,,
\label{Upsilonk}
\end{equation}
up to small (logarithmic) dependences on the coupling constant of the
$\chi$ field with the radiation fields. In Eq.~(\ref{Upsilonk}) we
have used comoving momentum $k \equiv |{\bf k}|$.
The above  expression (\ref{Upsilonk}) is the one we
used in our calculations for the bispectrum.
  

\end{document}